# Performance of TCP over ABR on ATM backbone and with various VBR traffic patterns[1]

Shiv Kalyanaraman, Raj Jain, Sonia Fahmy, Rohit Goyal and Jianping Jiang
The Ohio State University
Department of CIS
Columbus, OH 43210-1277
Email: {*shivkuma, jain, fahmy, goyal, jianping*}@cis.ohio-state.edu
and
Seong-Cheol Kim
Principal Engineer, Network Research Group
Communication Systems R&D Center
Samsung Electronics Co. Ltd.
Chung-Ang Newspaper Bldg.
8-2, Karak-Dong, Songpa-Ku
Seoul, Korea 138-160
Email: kimsc@metro.telecom.samsung.co.kr,

## Abstract

We extend our earlier studies of buffer requirements of TCP over ABR [10, 11, 12] in two directions. First, we study the performance of TCP over ABR in an ATM backbone. On the backbone, the TCP queues are at the edge router and not inside the ATM network. The router requires buffer equal to the sum of the receiver window sizes of the participating TCP connections. Second, we introduce various patterns of VBR background traffic. The VBR background introduces variance in the ABR capacity and the TCP traffic introduces variance in the ABR demand. Some simple switch schemes are unable to keep up with the combined effect of highly varying demands and highly varying ABR capacity. We present our experiences with refining the ERICA+ switch scheme to handle these conditions.

# 1   Introduction

With the proliference of multimedia traffic over the Internet, several technologies capable of handling such traffic efficiently are competing to replace various backbones and subnetworks of the Internet. The Asynchronous Transfer Mode (ATM) networks, which have been designed specifically to support integration of data, voice, and video applications is one of the key technologies in this competition.

ATM networks provide multiple classes of service. The Available Bit Rate (ABR) and the Unspecified Bit Rate (UBR) service classes have been developed specifically to support data applications. The ABR and UBR services differ in the way they control data traffic. The ABR service requires network switches to constantly monitor their load and feed the information back to the sources, which in turn dynamically adjust their input into the network [1, 6]. The UBR service does not provide any standard traffic management mechanism. However, the switches may monitor their queues and simply discard cells or packets of overloading users. It is interesting to see the performance of Internet (TCP) traffic over the ABR and UBR services. In this paper, we concentrate on TCP performance over ABR. TCP performance over UBR has also been studied in the literature [3, 4, 13, 5].

---

[1]Submitted to ICC'97, 8-12 June 1997, Montreal.
Available from http://www.cis.ohio-state.edu/~jain/papers/tcp_vbr.ps

One more reason for the study of TCP traffic over ABR is because TCP offers a unique workload for the performance analysis of ABR switch schemes. The ATM ABR traffic management schemes were initially analyzed using constant demand (infinite) sources and constant capacity (no VBR traffic) links. Recent analyses have given importance to the VBR traffic as a background workload because, the ATM link is in reality shared by multiple classes of traffic. The introduction of VBR background traffic makes the ABR capacity variable. The subsequent step was to find a better ABR demand model than the infinite traffic model. TCP traffic provides one such model.

TCP provides a reliable transfer of data using a window-based flow and error control algorithm [2]. When TCP runs over ABR, the TCP window-based control runs on top of the ABR rate-based control. The TCP traffic appears bursty (variable) at the ATM layer [11].

ABR switch schemes typically use the current demand and capacity to calculate feedback to the sources. The variable demand and variable capacity introduces variance in the measurements made by the switch schemes, and as a result, in the feedback given. It is interesting to study the effect of TCP (variable demand) and VBR (variable capacity) on ABR performance.

## 2   TCP Behavior over ABR

Though the application running on top of TCP is an infinite traffic application, the traffic seen at the ATM layer is bursty because of the TCP congestion avoidance mechanism [11]. The TCP behavior results in some effects seen by the ABR switch scheme:

**a)** Out-of-phase effect: No load or sources are seen in the forward direction while sources and RM cells are seen in the reverse direction.

**b)** Clustering effect: The cells from TCP connections come in clusters, and large intervals are required to sense the activity of multiple sources.

Due to these effects, switches may make errors in measuring quantities which they use to calculate feedback, for example, the load and activity of sources. These effects reduce as the network path gets completely filled by TCP traffic, and the ABR closed loop control becomes effective. The switch scheme then controls the rate of the sources.

Several authors have studied the performance of TCP over ABR and UBR services under lossy conditions [3, 4, 13, 5]. We show in [9] that, since TCP has a built in congestion avoidance mechanism, it does not lose too many cells under these conditions. The throughput is low because of the time lost during timeout and retransmission. Under these conditions a smaller TCP timer granularity and intelligent switch drop policies like Early Packet Discard (EPD) [14] help improve performance of TCP.

However, TCP achieves maximum throughput when there is no packet loss. We studied the ATM switch buffer requirements to allow zero-loss TCP transmission over ABR [10, 11, 12]. To achieve zero-loss, ABR service requires switch buffering which is only a small multiple of the round trip time and the feedback delay. The buffering depends heavily upon the switch scheme used.

Once the ATM source rates are controlled, the queues build up at the sources, and not at the switches. In effect, the ABR queues are pushed to the edge of the ATM network. In an ATM backbone network, the source is the edge router. In this paper, we quantify the buffering requirement at the edge router and discuss related issues.

The introduction of VBR traffic makes the ABR capacity variable resulting in more more variance in measurement. We examine the effect of using different VBR background patterns, the feedback delay and the switch scheme used. We present our experiences with refining the ERICA and ERICA+ switch schemes [7] to handle these conditions.

# 3    TCP Options And ERICA Parameters

We use a TCP maximum segment size (MSS) of 512 bytes. The MTU size used by IP is generally 9180 bytes and so there is no segmentation caused by IP. We implemented the window scaling option so that the throughput is not limited by path length. Without the window scaling option, the maximum window size is $2^{16}$ bytes or 64 kB. We use a maximum receiver window of 16x64 kB or 1024 kB. The network consists of three links of 1000 km max each and therefore, has a max one-way delay of 15 ms (or 291 kB at 155 Mbps). The maximum receiver window is, thus, greater than twice the one-way delay. We use a TCP timer granularity of 100ms. The timer is exercised only when there is packet loss.

The TCP data is encapsulated over ATM as follows. First, a set of headers and trailers are added to every TCP segment. We have 20 bytes of TCP header, 20 bytes of IP header, 8 bytes for the RFC1577 LLC/SNAP encapsulation, and 8 bytes of AAL5 information, a total of 56 bytes. Hence, every MSS of 512 bytes becomes 568 bytes of payload for transmission over ATM. This payload with padding requires 12 ATM cells of 48 data bytes each. Hence, the maximum receiver window of 1024 kB corresponds to 24576 cells over ATM.

In our simulations, we have not used the "fast retransmit and recovery" algorithms. The zero-loss buffer requirement is valid for fast retransmit and recovery too, since these algorithms are not exercised when there is zero-loss.

The ERICA algorithm [7] uses two key parameters: target utilization and averaging interval length. The algorithm measures the load and number of active sources over successive averaging intervals and tries to achieve a link utilization equal to the target. The averaging intervals end either after the specified length or after a specified number of cells have been received, whichever happens first. In the simulations reported here, the target utilization is set at 90%, and the averaging interval length defaults to 1 ms or 100 ABR input cells, represented as the tuple (1 ms, 100 cells). A modified version of the ERICA algorithm is used in the study of source end system queues. This version includes the averaging feature for the number of sources and a large averaging interval of (5 ms, 500 cells). The features and the reason for their choice is discussed in Section 5.2.4.

The ERICA+ algorithm is an extension of ERICA which uses the queueing delay as a additional metric to calculate the feedback. ERICA+ eliminates the target utilization parameter (set to 1.0) and uses four new parameters: a target queueing delay (T0 = 500 microseconds), two curve parameters (a = 1.15 and b = 1.05), and a factor which limits the amount of ABR capacity allocated to drain the queues (QDLF = 0.5).

## 4    The N Source + VBR Configuration

The "N Source + VBR" configuration shown in figure 1 has a single bottleneck link shared by the N ABR sources and possibly a VBR source. Each ABR source is a VBR source. All traffic is unidirectional. All links run at 155 Mbps. The links traversed by the connections are symmetric i.e., each link on the path has the same length for all the VCs. In our simulations, N is 15 and the link lengths may assume values 1000, 500, 100 and 1 km.

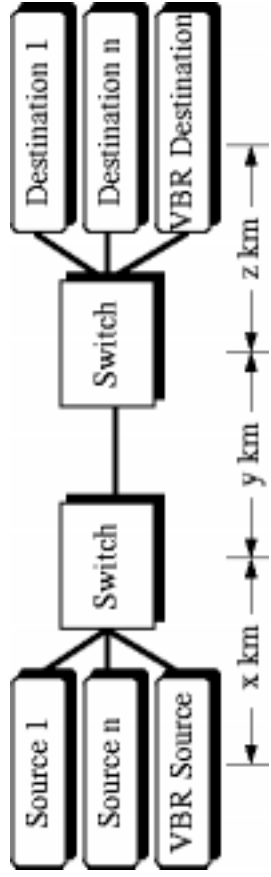

Figure 1: *n* Source + VBR Configuration

The individual link lengths determine the round trip time (RTT) and the feedback delay. Feedback delay is the sum of the delay for feedback from the switch to reach the source and the delay for the new load from the sources to reach the switch. It is at least twice the one-way propagation delay from the source to the switch. The feedback delay determines how quickly the feedback is conveyed to the sources and how quickly the new load is felt at the switch.

The VBR source when present is an ON-OFF source. The ON time and OFF time are defined in terms of a "duty cycle" and a "period". A pulse with a duty cycle of $d$ and period of $p$ has an ON time of $d \times p$ and and OFF time of $(1-d) \times p$. Our previous results of TCP over VBR used a duty cycle of 0.5 resulting in the ON time being equal to the OFF time. Unequal ON-OFF times used in this study cause new effects that were not seen before.

The VBR starts at $t = 2$ ms to avoid certain initialization problems. During the ON time, the VBR source operates at its maximum amplitude. The maximum amplitude of the VBR source is 124.41 Mbps (80% of link rate). VBR is given priority at the link, i.e, if there is a VBR cell, it is scheduled for output on the link before any waiting ABR cells are scheduled.

## 5    Summary of Results

First, we quantify the buffer requirement at the sources for zero-loss TCP transmission and discuss implications for backbone ATM networks. The VBR source is turned off in this study. We then introduce VBR traffic and examine the switch buffering requirement, studying the effect of varying the VBR ON-OFF periods, the ABR feedback delay and the ABR switch scheme.

## 5.1 TCP Performance over ATM Backbone Networks

The ATM source buffer requirement is derived by examining the maximum queues at the source when TCP runs over ABR. We also study the performance when sufficient buffers are not provided and discuss the implications for ATM backbone networks.

### 5.1.1 Source End System Queues in ABR

Table 1 shows the results with a 15-source configuration with link lengths of 1000 km (there is no VBR background). The link lengths yield a round trip time (propagation) of 30 ms and a feedback delay of 10 ms. We vary the size of the source end system buffers from 100 cells to 100000 cells per VC (second column). These values are compared to the maximum receiver window size (indicated as "Win" in the table) which is 1024 kB = 24576 cells. The switch has infinite buffers and uses a modified version of the ERICA algorithm [7] including the averaging feature for the number of sources and an averaging interval of (5 ms, 500 cells) as described in Section 5.2.4.

The maximum source queue values (third column) are tabulated for every VC, while the maximum switch queue values (fourth column) are for all the VCs together. When there is no overflow the maximum source queue (third column) measured in units of cells is also presented as a fraction of the maximum receiver window. The switch queues are presented as a fraction of the round trip time (indicated as "RTT" in the table). The round trip time for this configuration is 30 ms which corresponds to a "cell length" of 30 ms × 368 cells/ms = 11040 cells.

The last column tabulates the aggregate TCP throughput. The maximum possible TCP throughput in our configuration is approximately: 155.52 × (0.9 for ERICA Target Utilization) × (48/53 for ATM payload) × (512/568 for protocol headers) × (31/32 for ABR RM cell overhead) = 110.9 Mbps.

Table 1: Source Queues in ABR

| # | Source Buffer (cells) | Max Source Q (cells) | Max Switch Q (cells) | Total Throughput |
|---|---|---|---|---|
| 1. | 100 (< Win) | > 100 (overflow) | 8624 (0.78×RTT) | 73.27 Mbps |
| 2. | 1000 (< Win) | > 1000 (overflow) | 17171 (1.56×RTT) | 83.79 Mbps |
| 3. | 10000 (< Win) | > 10000 (overflow) | 17171 (1.56×RTT) | 95.48 Mbps |
| 4. | 100000 (> Win) | 23901 (0.97×Win) | 17171 (1.56×RTT) | 110.90 Mbps |

In rows 1, 2 and 3 of Table 1, the source has insufficient buffers. The maximum per-source queue is equal to the source buffer size. The buffers overflow at the source and cells are dropped. TCP then times out and retransmits the lost data.

TCP performance under these conditions (of insufficient source buffers and sufficient switch buffers) is similar to its performance when the switch has insufficient buffers and the source has sufficient buffers [9]. This is briefly described in section 1.

Also observe that the switch queue reaches its maximum possible value for this configuration (1.56×RTT) given a minimum amount of per-source buffering (1000 cells = 0.04×Win). The switch buffering requirement is under 3×RTT as predicted in [10, 11, 12].

The sources however require one receiver window's worth of buffering per VC to avoid cell loss. This hypothesis is substantiated by row 4 of Table 1 which shows that the maximum per-source queue is 23901 cells = 0.97×Win. The remaining cells (0.03×Win) are traversing the links inside the ATM network. The switch queues are zero because the sources are rate-limited by the ABR mechanism [9]. The TCP throughput (110.9 Mbps) is the maximum possible given this configuration, scheme and parameters.

The total buffering required for N sources is the sum of the N receiver windows. Note that this is the same as the switch buffer requirement for UBR [5]. In other words, the ABR and UBR services differ in whether the sum of the receiver windows' worth of queues is seen at the source or at the switch.

### 5.1.2   Implications for ATM Backbone Networks

If the ABR service is used end-to-end, then the TCP source and destination are directly connected to the ATM network. The source can directly flow control the TCP source. As a result, the TCP data stays in the disk and is not queued in the end-system buffers. In such cases, the end-system need not allocate large buffers. ABR is better than UBR in these (end-to-end) configurations since it allows TCP to scale well.

However, if the ABR service is used on a backbone ATM network, the end-systems are edge routers which are not directly connected to TCP sources. These edge routers may not be able to flow control the TCP sources except by dropping cells. To avoid cell loss, these routers need to provide one receiver window's worth of buffering per TCP connection. The buffering is independent of whether the TCP connections are multiplexed over a smaller number of VCs or they have a VC per connection. For UBR, these buffers need to be provided inside the ATM network, while for ABR they need to be provided at the edge router. If there are insufficient buffers, cell loss occurs and TCP performance degrades.

The fact that the ABR service pushes the congestion to the edges of the ATM network while UBR service pushes it inside is an important benifit of ABR for the service providers. In general, UBR service requires more buffering in the switches than the ABR service.

## 5.2   Performance of TCP over ABR with VBR Background

We now continue our study of ABR switch buffering by introducing VBR traffic in addition to the 15 ABR sources. All link lengths are 1000km. The round trip time is 30 ms and the feedback delay is 10 ms.

We use the ERICA+ algorithm [7] in our results. The ERICA+ algorithm maximizes the efficiency by allowing 100% utilization in the steady state. It also tries to achieve a target queueing delay. In the presence of VBR, the target is never achieved due to the variance in capacity. However, since the ABR capacity is scaled as a function of queue length, the queue maximum can be controlled.

Although we had invented ERICA+ to allow 100% utilization of expensive links, we found that ERICA+ is helpful in controlling queues and stability in cases with high variance in demand or capacity.

Table 2 shows the results of a 3x3 full-factorial experimental design [8] used to identify the problem space with VBR background traffic. We vary the two VBR model parameters: the duty cycle (d) and the period (p). Recall that, with parameters d and p, the VBR ON time is d×p and the VBR OFF time is d×(1-p). Each parameter assumes three values. The duty cycle assumes values 0.95, 0.8 and 0.7 while the period may be 100 ms (large), 10 ms (medium) and 1 ms (small).

The maximum switch queue is also expressed as a fraction of the round trip time (30 ms = 30 ms × 368 cells/ms = 11040 cells).

Table 2: Effect of VBR ON-OFF Times

| # | Duty Cycle(d) | Period (p) | Max Switch Q |
|---|---|---|---|
| | (ms) | (cells) | |
| 1. | 0.95 | 100 | 2588 (0.23×RTT) |
| 2. | 0.8 | 100 | 5217 (0.47×RTT) |
| 3 | 0.7 | 100 | 5688 (0.52×RTT) |
| 4. | 0.95 | 10 | 2709 (0.25×RTT) |
| 5. | 0.8 | 10 | DIVERGENT |
| 6. | 0.7 | 10 | DIVERGENT |
| 7. | 0.95 | 1 | 2589 (0.23×RTT) |
| 8. | 0.8 | 1 | 4077 (0.37×RTT) |
| 9. | 0.7 | 1 | 2928 (0.26×RTT) |

### 5.2.1   Effect of VBR ON-OFF Times

Rows 1,2 and 3 of Table 2 characterize large ON-OFF times (low frequency VBR). Observe that the (maximum) queues are small fractions of the round trip time. The queues which build up during the ON times are drained out during the OFF times. Given these conditions, VBR may add at most one RTT worth of queues. ERICA+ further controls the queues to small values.

Rows 4,5 and 6 of Table 2 characterize medium ON-OFF times. We observe that rows 5 and 6 have divergent (unbounded) queues. The effect of the ON-OFF time on the divergence is explained as follows. During the OFF time the switch experiences underload and may allocate high rates to sources. The duration of the OFF time determines how long such high rate feedback is given to sources. In the worst case, the ABR load is maximum whenever the VBR source is ON to create the largest backlogs.

On the other hand, the VBR OFF times also allow the ABR queues to be drained out, since the switch is underloaded during these times. Larger OFF times may allow the queues to be completely drained before the next ON time. The queues will grow unboundedly (i.e., diverge) if the queue backlogs accumulated after ON and OFF times never get cleared.

Rows 7,8 and 9 of Table 2 characterize small ON-OFF times. Observe again that the queues are small fractions of the round trip time. Since the OFF times are small, the switch does not have enough time to allocate high rates. Since the ON times are small, the queues do not build up significantly in one ON-OFF cycle. On the other hand, the frequency of the VBR is high. This means that the VBR changes much faster than the time required for sources to respond to feedback. ERICA+ however controls the queues to small values in these cases.

### 5.2.2    Effect of Feedback Delays

Another factor which interacts with the VBR ON-OFF periods is the feedback delay. We saw that one of the reasons for the divergent queues was that switches could allocate high rates during the VBR OFF times. The feedback delay is important in two ways. First, the time for which the switch may allocate high rates is the minimum of the feedback delay and the VBR OFF-time. This is because, the load due to the high rate feedback is seen at the switch within one feedback delay. Second, when the load due to the high rate feedback is seen at the switch, it takes at least one feedback delay to reduce the rates of the sources.

The experiments shown in Table 2 have a long feedback delay (10 ms). The long feedback delay allows the switch to allocate high rates for the entire duration of the VBR OFF time. Further, when the switch is overloaded, the sources takes 10 ms to respond to new feedback. Therefore, given appropriate value of the ON-OFF times (like in rows 4,5 of Table 2), the queues may diverge.

Table 3 shows the effect of varying the feedback delay and round trip time. We select the divergent case (row 4) from Table 2 and vary the feedback delay and round trip time of the configuration.

Table 3: Effect of Feedback Delay

| # | Feedback Delay(ms) | RTT (ms) | Duty Cycle (d) | Period (p) (ms) | Max Switch Q (cells) |
|---|---|---|---|---|---|
| 1. | 1 ms | 3 ms | 0.8 | 10 ms | 4176 (0.4×RTT) |
| 2. | 5 ms | 15 ms | 0.8 | 10 ms | DIVERGES |
| 3. | 10 ms | 30 ms | 0.8 | 10 ms | DIVERGES |

Row 1 in Table 3 shows that the queues are small when the feedback delay is 1 ms (metropolitan area network configuration). In fact, the queues will be small when the feedback delay is smaller than 1 ms (LAN configurations). In such configurations, the minimum of the OFF time (2 ms) and the feedback delay (< 1 ms) is the feedback delay. Hence, in any VBR OFF time, the switch cannot allocate high rates to sources long enough to cause queue backlogs. The new load is quickly felt at the switch and feedback is given to the sources.

Rows 2 and 3 in Table 3 have a feedback delay longer than the OFF time. This is one of the factors causing the divergence in the queues of these rows.

### 5.2.3   Effect of Switch Scheme

The TCP traffic makes the ABR demand variable. The VBR background makes the ABR capacity variable. In the presence of TCP and VBR, the measurements used by switch schemes are affected by the variance. The errors in the metrics are reflected in the feedback. The errors in the feedback result in queues. Switch schemes need to be robust to perform under such error-prone conditions. Another effect of errors is that the boundary conditions of the scheme are encountered often. The scheme must be designed to handle such conditions gracefully. We study the robustness issues in ERICA and make adjustments needed to reduce the effect of the variance.

As an example, consider the case when the VBR ON-OFF periods are very small (1 ms ON, 1 ms OFF). The resulting variance can be controlled by a switch scheme like ERICA+ which uses the queueing delay to calculate feedback (in addition to input rate etc). The basic ERICA algorithm without queue control cannot handle this level of variance.

The ERICA+ algorithm uses the queue length as a secondary metric to reduce the high allocation of rates. However, ERICA+ has a limit on how much it can reduce the allocation. Given sufficient variance, the limit can be reached. This means that even the minimum rate allocation by ERICA+ causes the queues to diverge. This reason, along with the discussion on ON-OFF times and feedback delays explains the divergent cases in Tables 2 and 3.

### 5.2.4   Reducing the Variance In ERICA+

We tackle these problems by reducing the effect of variance on the scheme measurements in three ways:

1. First, one way to reduce variance in measurements is to measure over longer intervals. Longer intervals yield averages which have less variance. However, making the intervals too long increases the response time, and queues may build up in the interim.

2. Second, we average the measurements over several successive intervals. The ERICA scheme uses two important measurements: the overload factor (z) which is the ratio of the input rate and the target ABR rate, and the number of active sources (N). We re-examine how the scheme depends on these metrics and design an appropriate averaging technique for each of them.

   - The overload factor (z) is used to divide the current cell rate of the source to give what we call the "VC share". The VC share is one of the rates which may be given as feedback to the source. If the overload factor (z) is underestimated, the VC share increases. The overload factor is usually not overestimated. However, if the interval length is small, the estimated values may have high variance.

The overload factor (z) can suddenly change in an interval if the load or capacity in that interval changes due to the variance. The out-of-phase effect of TCP may lead to no cells being seen in the forward direction (z = 0, a huge underestimate !), while BRM cells are seen in the reverse direction. The switch will then allocate a high rate in the feedback it writes to the BRM cell.

We have designed two averaging schemes for the overload described in reference [7]. Both schemes use an averaging parameter "$\alpha_z$". The first scheme is similar to the exponential averaging technique for a random variable. However, it differs because it resets the averaging mechanism whenever the instantaneous value of overload is measured to be zero or infinity. The second scheme does not ignore the outlier values (zero or infinity) of the overload factor. Further, it averages the overload by separately averaging the input rate and capacity, and then taking the ratio of the averages. It can be shown [8] that this is theoretically the right way to average a ratio quantity like overload.

- The number of active sources (Na) is used to calculate a minimum fairshare that is given to any source. If Na is underestimated, then the minimum fairshare is high leading to overallocation. If Na is overestimated, then the minimum fairshare is low. This may result in slower transient response, but does not result in overallocation.

  The number of active sources can fluctuate if some sources are not seen in an interval. Further, due to the clustering effect of TCP, cells from just a few VCs may be seen in an interval leading to an underestimate of Na.

  In averaging Na, the scheme maintains an activity level for each source. The activity level of the source is set to one when any cell of the source is seen in the interval. However, when no cell from a source is seen in an interval, the scheme "decays" the activity level of the source by a factor, "$\alpha_n$". Hence, the source becomes inactive only after many intervals. A recommended value of $\alpha_n$ is 0.9. Roughly, the Na measured with this value of $\alpha_n$ is approximately equal to the Na measured without averaging over an averaging interval 8 or 9 times larger than the current averaging interval.

3. Third, we modify the response to boundary conditions of the scheme. This allows the scheme to handle the boundary conditions gracefully. Specifically, the number of active sources is set to one if it is measured to be below one. The second method of overload factor averaging does not allow the overload factor be zero or infinity. However, outlier measurements are not ignored in the averaging method.

The ERICA+ scheme with these modifications controls the ABR queues without overly compromising on TCP throughput. Table 4 shows the results of representative experiments using these features.

Row 1 shows the performance with the averaging of Na and z turned on on a formerly divergent case. Observe that the queue converges and is small. The parameter $\alpha_z$ is 0.2, which is roughly equivalent to increasing the averaging interval length by a factor of 5. Hence, we try the value (500 cells, 5 ms) as the averaging interval length, without the averaging of overload factor. Row 2 shows that the queue for this case also converges and is small.

Table 4: Effect of Switch Scheme

| # | Averaging Interval (T ms, n cells) | Averaging of Na on ? ($\alpha_n = 0.9$) | Averaging of z on ? ($\alpha_z = 0.2$) | d | p(ms) | Max Switch Queue (cells) |
|---|---|---|---|---|---|---|
| 1. | (1,100) | YES | YES | 0.7 | 20 | 5223 |
| 2. | (5,500) | YES | NO | 0.7 | 20 | 5637 |

# 6    Summary

We have presented further results on the issue of buffering requirements for TCP over ABR. The first result deals with the source end system queues, and has significance in ABR backbone configurations. Though the ABR switch buffering requirement is small, the ATM source buffering required is equal to the sum of the TCP receiver window sizes. This is the buffering required in edge routers of the ATM network.

We then carefully study the impact of VBR background traffic on switch buffering. We find that the ON-OFF times, the feedback delays, and a switch scheme sensitive to variance in ABR load and capacity may combine to create worst case conditions where the ABR queues diverge. We enhance the ERICA+ scheme in three ways to reduce the effect of the variance and allow the convergence of the ABR queues, without compromising on the efficiency.

---

[2] Throughout this section, AF-TM refers to ATM Forum Traffic Management sub-working group contributions.

---

[3] All our papers and ATM Forum contributions are available through http://www.cis.ohio-state.edu/~jain